\documentclass[aps,floatfix,showpacs,epsfig]{revtex4}
\usepackage{epsf}
\usepackage{graphicx}
\usepackage{amssymb}
\usepackage{amssymb,latexsym}
\usepackage{amsmath}
\usepackage{mathrsfs}
\usepackage[mathscr]{eucal}
\usepackage{color}
\usepackage{natbib}
\def\be{\begin{equation}}
\def\ee{\end{equation}}
\def\ba{\begin{eqnarray}}
\def\ea{\end{eqnarray}}
\def\Lie{{\cal L}}
\begin{document}
\title{Cosmic Matter Flux May Turn Hawking Radiation Off}

 \author{Javad T. Firouzjaee${}^{1,2}$}
\affiliation{${}^1$School of Astronomy, Institute for Research in Fundamental Sciences (IPM), P. O. Box 19395-5531, Tehran, Iran }
 \email{j.taghizadeh.f@ipm.ir}
\author{George F R Ellis${}^{2}$}
\affiliation{${}^2$Mathematics and Applied Mathematics Department, University of Cape Town, Rondebosch, Cape Town 7701, South Africa}
 \email{gfrellisf@gmail.com}

\begin{abstract}
\textbf{Abstract:} An astrophysical (cosmological) black hole forming in a cosmological context will be subject to a  flux of infalling matter and radiation, which will cause the outer apparent horizon (a marginal trapping surface) to be spacelike \cite{ellisetal14}. As a consequence the radiation emitted close to the apparent horizon no longer arrives at infinity with a diverging redshift. Standard calculations of the emission of Hawking radiation then indicate that no blackbody radiation is emitted to infinity by the black hole in these circumstances, hence there will also then be no black hole evaporation process due to emission of such radiation as long as the matter flux is significant. The essential adiabatic condition (eikonal approximation) for black hole radiation gives a strong limit to the black holes that can emit Hawking radiation.  We give the mass range for the black holes that can radiate, according to their cosmological redshift, for the special case of the cosmic blackbody radiation (CBR) influx (which exists everywhere in the universe). At a very late stage of black hole formation when the CBR influx decays away, the black hole horizon becomes first a slowly evolving horizon and then an isolated horizon; at that stage, black hole radiation will start. This study suggests that the primordial black hole evaporation scenario should be revised to take these considerations into account.
\end{abstract}
%
%

\maketitle

\tableofcontents
\section{Introduction}

This paper considers formation of an astrophysical black hole in a cosmological context \cite{cosmological black hole}. Unlike the simple Schwarzchild or Kerr black holes, which are static or stationary respectively, cosmological black holes have dynamic behaviour and are surrounded by different kinds of matter and radiation in a changing background. In the case where the geometry is non-static, the need for a local definition of black holes and their horizons has led to concepts such as an isolated horizon \cite{isolated}, Ashtekar and Krishnan's dynamical horizon (DH) \cite{ashtekar02}, and Booth and Fairhurst's slowly evolving horizon \cite{boothslow}.
Although the different aspects of the dynamical nature of the astrophysical black hole  \cite{flux-bh} and primordial black hole  \cite{flux-pbh} have been studied, there have not been serious attempts to consider black hole particle creation in the vicinity of dynamical horizons.\footnote{It has been commented that horizons (whether apparent horizons or event horizons) do not somehow ``emitted'' particles/radiation because quantum field theory in curved
 spacetime does not describe the Hawking process in this manner at all. But in our view this is just semantics.  There are papers that show that the radiation appears to be emitted from the region of space-time near the apparent horizon, which this seems a good enough justification for saying ``it is emitted from the vicinity of the apparent horizon''.}\\

 It was shown in a previous paper \cite{ellisetal14} that in a realistic cosmological situation, an infalling energy density such as incoming cosmic blackbody radiation (CBR) would make the outer apparent horizon (the Outer Marginally Trapped 3-Surface, or OMOTS) become spacelike.  When the apparent horizon is spacelike, it seems likely that some emitted radiation will fall into the singularity, and some will escape to infinity. This is because the radiation is emitted in the vicinity of the apparent horizon, so some may be emitted from far enough away that it will escape the event horizon. One might call this  part of the radiation that escapes to infinity ``Hawking radiation''; that is a question of terminology. We rather refer to all  radiation from the vicinity of the apparent horizon as Hawking radiation, whether it reaches infinity or not.\\

Hawking's quantum field theory approach to black hole radiation \cite{Haw73, wald-14, Haw74}, which applies to late time stationary black
holes, is not a suitable method for calculating the Hawking temperature in the case of a fully 
dynamical black hole, where one has to solve the field equations in a changing background. Even the semiclassical stress tensor approach \cite{BirDav84} has limitations related to finding the effective action and taking into account the backscattering problem. In such cases, one should look for alternative approaches allowing one to calculate the Hawking radiation in a dynamical context \cite{parikh2000, visser03, tunnelingbh}. This radiation is plausibly emitted from the vicinity of apparent horizons rather than from near  the global event horizon \cite{visser10}.\\

The tunneling picture of Hawking radiation emission \cite{tunnelingbh,parikh2000} suggests that when there is infalling matter or radiation, there will be no emission of blackbody radiation arriving at future infinity from the  vicinity of the OMOTS surface, because it is spacelike  \cite{ellis13,ellisetal14}. Additionally Hawking radiation emitted from the inner apparent horizon (the Inner Marginally Trapped 3-Surface, or IMOTS) would have been trapped, so on this view, no Hawking radiation would escape to infinity \cite{ellis13}. However this is based on a particle picture that can be queried.\\

Here we show that if one uses the eikonal approximation \cite{visser03}, that result is confirmed: when an infalling matter or radiation flux occurs, no Hawking radiation is emitted from the spacelike OMOTS surface. However additionally, because of the way the phase relates to redshift, the IMOTS surface will also not emit Hawking Radiation.  In other words, both tunneling arguments (WKB approximation) \cite{parikh2000} and eikonal arguments \cite{visser10} suggest that a matter or CBR radiation flux turns off Hawking radiation to infinity, implying that no black hole explosions \cite{Haw74} would occur when black holes form in a cosmological context as long as the matter flux is significant. However at very late times, all the matter that can do so will have fallen into the black hole in the $\Lambda CDM$ background and the CBR will have decayed away to almost zero. An isolated event  horizon will then form and, on the eikonal view, will then lead to the standard picture of black hole radiation. \\

One should note here that while the mathematically clearest derivation of the Hawking effect in the usual non-dynamical context involves calculating the evolution of the two-point function of the quantum field backward in time from infinity to near the horizon, the relevant properties of the calculation in that case can also be seen clearly by using the different method given in Hawking's original paper. However neither method is appropriate in the
 context we consider, where the semi-classical radiation  we are considering may not escape to infinity at all. That is why we use the methods we do, which are suitable for a dynamical context. Dynamical black holes may or may not emit also Hawking radiation. This is a valid question, and  we should not be using any framework that doesn't allow it to be answered because it assumes the geometry is close to Schwarzschild. The detailed discussion in  most papers deriving Hawking radiation, for example \cite{wald-14}, are confined to the case of non-dynamical black holes, and we are considering the more realistic dynamical case. It has been suggested that the deviations from Schwarzschild due to infalling matter would have a completely negligible effect; however this is not supported by work on modelling dynamical black holes \cite{flux-bh,flux-pbh}.
 \\

 We also consider Hawking's original argument \cite{Haw73} using Bogoliubov coefficients for deriving the black hole thermal radiation for stationary black holes (when there is no influx of matter or radiation). This argument relates radiation emission to the way the geometry of outgoing null geodesics leads to exponential rescaling of null geodesic generators of past null infinity and future null infinity \cite{visser10}.
However when there is infalling matter or radiation, the fact that the OMOTS surface is spacelike causes the event horizon to shift outwards relative to when there is no infalling energy density, and the spacetime  geometry is then such that the relevant outgoing null geodesics never reach infinity  \cite{ellisetal14}: rather they fall into the future spacelike singularity (see Fig.(\ref{bh_Ellis_javad_1})). The Hawking argument does not then apply. \\

\textbf{Event horizon and backreaction problem:}
Generally, we can divide black hole formation and evolution into three steps. \\

The first steps involves the gravitational collapse of matter. During this step,  apparent horizons and the singularity form, and we can define the classical event horizon. Most of the gravitational physics can be described by classical physics.  During this step, we can study the nature of the horizons (IMOTS and OMOTS) and their location, shell crossings and matter flux rate, and so on. As discussed in this paper, the eikonal (adiabatic) approximation  will not be valid  during this step.\\

The second step is when the black hole becomes quasi-static, and we can apply the adiabatic condition to it. The black hole starts to radiate and lose mass, and we have to consider semiclassical physics in order to examine the time evolution \cite{visser07}. During this step determining the location of the event horizon is complicated, because in order to find the event horizon location we need to know the whole story of the gravitational collapse for all time. To do this, we need to solve the backreaction problem, but it seems that no one has a comprehensive theory to do so. Since the black hole at this step emits quanta of much smaller energy than the whole mass, the adiabatic approximation used in the
quantum calculations of the emission will be valid.\\

The final step which determines the fate of the black hole is the quantum gravitational  step. It seems that we can only examine the fundamental question of black hole information and the event horizon during this step. \\

Our work here considers the first step -- the black hole evolution -- and investigates when it reaches the second step.
This argument does not cover the back reaction problem and possible eventual black hole evaporation. To consider this, one will need to check what the effective stress-tensor associated with particle production is, and also that radiation emission from the central collapsing fluid \cite{BirDav84} does not prevent formation of apparent horizons and the associated singularities \cite{HawEll73}.  These issues will be the topic of further papers. However the tunneling approach is used by many, as is the eikonal approximation: this paper takes the argument of \cite{ellisetal14} forward by showing these two approaches agree that existence of the matter flux turns off Hawking radiation and so suppresses black hole explosions.  \\

\textbf{This paper:}  We will show the close relation between the  eikonal approximation and particle tunneling picture in section II.
In Section III, we discuss the eikonal approximation relation to the redshift of null geodesics. Section IV considers  this approximation for some cosmological black hole models. In Section V, we consider the matter and radiation flux and its evolution in time. Section VI looks at how the exponential relation between null geodesic parameters on future and past null infinities does not occur when the OMOTs is spacelike. We then conclude in section VII.
In carrying out this study, it is convenient to think of three successive approximations. First we consider the standard case \cite{Haw73} with no matter flux effect and a static exterior spacetime. Second, we consider black holes with a constant positive influx of matter or radiation at all times \cite{ellisetal14}. Finally, we consider the realistic case where the matter flux rate is non zero at all times, but is decaying away to zero in the late time expanding universe.

\section{The Eikonal approximation and the tunneling picture}
In general relativity, the classical field $\phi$ without potential solves the massless Klein-Gordon equation
\begin{equation}\label{eq:KG}
g^{\mu \nu} \nabla 
_\mu \nabla
_\nu \phi=0.
\end{equation}
To solve it we can locally use the \textit{eikonal approximation}
\be\label{eik111}
\phi = a e^{i \psi}
\ee
where the amplitude $a(x^i)$ varies much more slowly than the phase $\psi(x^i)$ when
 the phase $\psi$ is rapidly varying: $\psi \gg 1$. The eikonal approximation can be presented according to the adiabatic condition which will be discussed in Appendix~\ref{C}.
Then (\ref{eq:KG}) gives the \textit{eikonal equation}
\be
g^{\mu \nu} \nabla 
_\mu \psi \nabla 
_\nu \psi= g^{\mu \nu} \partial
_\mu \psi  \partial
_\nu \psi=0 \label{eikonal}
\ee
for $\psi$.  In analogy with a wave in Minkowski space time, $k_i = \partial_i \psi $ is the wave vector (here Latin indexes run from 1 to 3) and $w= - \frac{\partial \psi}{\partial t}$ the frequency of the wave measured in the coordinate frame. For a  preferred observer with four-velocity $u^\mu$, the frequency measured by the observer is $ w = - u^\mu \partial_\mu \psi $. \\

We consider the case of spherical waves: $a=a(r,t)$, $\psi = \psi(r,t)$.
Over a small region of space time for a local observer the eikonal $\psi$ can then be expanded to  first order as
\be
\psi = \psi_0 +\frac{\partial \psi}{\partial t} t + \frac{\partial \psi}{\partial r} r
\ee
For a  stationary space time the eikonal can be written
\begin{equation}\label{eik11}
\psi = \psi_0 +w t \pm \int^r k(r') dr'
\end{equation}
where the +(-) sign in front of the integral indicates that the radial wave is ingoing (outgoing) and $k(r):=k_r$.  \\

The geometrical optics corresponds to the limiting case of small wavelength $\lambda \rightarrow 0$ which satisfies the eikonal equation (\ref{eikonal}).  Then the eikonal equation (\ref{eikonal}) is like the Hamilton-Jacobi equation
\be
g^{\mu \nu} \partial_\mu S \partial_\nu S=0
\ee
where the action $S$ is related to the momentum by $p_i = \frac{\partial S}{\partial x_i}$ and the Hamiltonian is  $H= - \frac{\partial S}{\partial t}$.  Comparing these formula with the field case, we see that wave vector plays the same role as momentum of the particle and frequency plays the role of the Hamiltonian or energy of the particle in  geometrical optics:
\begin{equation}
k_i \Leftrightarrow p_i,\,\,\, w \Leftrightarrow E \simeq H.
\end{equation}

To consider  Hawking radiation for a dynamical black hole in the spherically  symmetric case in Painlev$\acute{e}$-Gullstrand coordinates, Visser \cite{visser03} has used this ansatz  for the $s$ wave:
\be
\phi =A(r,t) exp[ \mp i (w t - \int^r k(r') dr')] \label{vissere}
\ee
which is basically the same as using (\ref{eik11}) in (\ref{eik111}). This form of the wave is valid provided the geometry is slowly evolving on the timescale
of the wave. \\

On the other hand, Hartle and Hawking \cite{hawking76} obtained particle production in stationary 
black hole
space-times using a semi-classical analysis which does not require knowledge of
the wave modes. This method has been extended to different Schwarzchild coordinates  in  \cite{pady2000}.  In their method, the  ratio between emission and absorption
probabilities is given by
\be
\Gamma \sim e^{- \beta w} =\frac{P_{[emission]}}{P_{[absorption]}},
\ee
where the probability $P$ is the square of the amplitude of the field: $P = |\phi|^2$.
By inserting  (\ref{vissere}) in this equation we get
\be
\Gamma \sim e^{- 2 Im \int^r k(r') dr'} .
\ee
 The term $\int^r k(r') dr'$ has a pole singularity on the apparent horizon  and gives an imaginary part in this coordinate system. If other coordinates had been chosen, one should have considered the temporal contribution to  the emission rate \cite{akhmedov}.  If one wants to use the particle picture for the wave and take $k(r) = p(r) $ then we will get the emission rate in the particle picture \cite{parikh2000}
\be
\Gamma \sim e^{- 2 Im \int^r p(r) dr} = e^{- 2 Im S}
\ee
where $S$ is the particle action which satisfies the Hamilton-Jacobi equation and for the Painlev$\acute{e}$ coordinate system we have
\begin{equation}
Im \int^r p(r) dr = Im S.
\end{equation}
As a result, the Visser eikonal method \cite{visser03}
and the Parikh  and Wilczek tunneling method \cite{parikh2000} are similar although they have different wave and particle pictures. However there is one key difference: the tunneling method  cannot apply for particle production whenever the MOTS surface is spacelike, because the whole concept of tunneling only  makes sense for a timelike surface (where `inside' and `outside' are well defined concepts); however the eikonal method (which is based on a wave rather than particle picture) can give particle production when the surface is spacelike and there is  slow evolution of the geometry. This difference will be important in the sequel.     \\

\textbf{Role of the eikonal approximation:}
In contrast to Minkowski space time, in which the definition of a particle  with momentum $k$ is based on a decomposition of
fields into plane waves $e^{i(wt-kx)}$, in the dynamic case the spatial size of the wave (particle) packet varies with the dynamics of the space time, and a plane wave cannot describe it. In spacetimes with a slowly changing geometry, the so-called adiabatic vacuum allows defining a meaningful notion of particles that can be applied for an evolving space time \cite{parker69}. The adiabatic vacuum prescription relies on the WKB (eikonal) approximation for solutions of the wave equation. As stressed by Barcelo et.al \cite{visser10}, physically the adiabaticity constraint (eikonal approximation) is equivalent to the statement that a photon emitted near the peak of the Planckian spectrum should not see a large fractional change in the peak energy of the spectrum over one oscillation of the electromagnetic field (that is, the change in spacetime geometry is adiabatic as seen by a photon near the peak of the Hawking spectrum).  The eikonal approximation for having particle creation may have a more profound meaning if we examine the quantum stress-tensor.

\section{The Geometric optics approximation, the wave phase, and redshift}
In the vacuum (Schwarzschild) case, the event horizon is associated with infinite redshift in the following sense: if a freely falling object drops into the black hole,  as it approaches the event horizon, an external observer will see it with ever increasing redshift; as it crosses the event horizon, the redshift diverges \cite{HawEll73}.  Anywhere that Hawking radiation is associated with an event horizon \cite{Haw73},  it is associated with an infinite redshift surface.\\

In this section we show that that association is not a coincidence: it is essential to the radiation process, and remains true even in the case of a dynamical horizon. \\

Consider the eikonal equation for the wave $\varphi=A(t,r)e^{\pm i \psi}$.  It is known that  $k_\mu=\nabla_\mu \psi$ , the normal vector of the constant phase plane, describes wave propagation, and is a null vector:
\begin{equation}\label{eq:null}
k_\mu k^\mu =0.
\end{equation}
 Differentiating this equation gives $$ k_\mu \nabla_\nu k^\mu =0.$$ Since $\nabla_\nu k_\mu=\nabla_\nu\nabla_\mu \psi=\nabla_\mu\nabla_\nu \psi=\nabla_\mu k_\nu$ we get the geodesic equation
\be
k^\nu \nabla_\nu k_\mu=0.
\ee
 In other words, the null geodesic vector derived from the eikonal equation is affinely parametrized, and
equation (\ref{eq:null}) is the same as the eikonal equation (\ref{eikonal}).\\

Let's look at light propagation from an emitter ($e$) to an observation point ($o$). The frequency which is measured by a
observer with 4-velocity $\tilde{u}^\mu=\frac{dx^\mu}{d\lambda}$ is
$w=k_\mu \tilde{u}^\mu$ ($\lambda$ is proper time for the time like
observer).
 Thus, the redshift at the observer point is
\begin{equation}
1+z=\frac{\nu_e}{\nu_o}=\frac{(g_{\mu\nu}k^\mu \tilde{u}^\nu
)_e}{(g_{\mu\nu}k^\mu \tilde{u}^\nu )_o}.
\label{redshift}\end{equation}
Note that  $k^\mu$ must be an affinely parametrized null vector. An infinite redshift surface is a surface such  that the redshift of light
arriving at that surface becomes infinite: $(1+z)\rightarrow \infty$. The geometric optics  approximation is satisfied near
an infinite redshift surface.\\

 Let us expand the wave phase near the eikonal approximation case $\psi \gg 1$:
\ba
\psi - \psi_0 &=& \nabla_\mu \psi dx^\mu = k_\mu
\frac{dx^\mu}{d\lambda}d\lambda\nonumber\\
&=& \nu d\lambda= (1+z)\nu_0 d\lambda.
\ea
In the second line we have used the redshift equation (\ref{redshift}). This equation shows that the phase of the wave near the infinite redshift surface is
very big: $\psi \gg 1$.\\

According to the discussion by Visser \cite{visser03}, the eikonal (geometric optics) approximation is an essential feature for occurrence of  black hole radiation. Therefore, having geometric optics valid 
in the close vicinity of the apparent horizon is a necessary condition for demonstrating  existence of Hawking radiation by the  
eikonal  method. We now see that we can use existence of an infinite  or very large redshift surface as a criterion for when this is satisfied. \\

This is actually associated with the exponential piling up of the null affine parameter relationship on future null infinity that is often seen as the key to existence of Hawking radiation (\cite{Haw73}  and see Section 6).
The reason is as follows: whenever there is a timelike Killing vector field $\xi_a$, the energy $E$ of a photon is given by $E = - \xi_a k^a$.
On a bifurcate Killing horizon, where a Killing vector field changes from timelike to spacelike \cite{Boy69}, the Killing vector parameter $\xi$: $\xi^a = dx^a/d\xi$  and the geodesic affine parameter $v$: $k^a = dx^a/dv$ are related exponentially: $v = \exp(\kappa \xi)$, $\kappa \neq 0$, which leads to the affine parameter relationship between past and future null infinity discussed in \cite{Haw73}. It follows that  \begin{equation}\label{eq:param}
k^a= \exp (-\kappa v) \xi^a,\,\ 
\end{equation} is parallely propagated on the null horizon \cite{Boy69}, which  leads to the divergent redshift relation as a null geodesic parallel to the horizon approaches the horizon. Note that there  will be no such divergence in the case when the Killing vector field does not change from timelike to spacelike on the horizon, but rather is null in an open neighbourhood. This is the case when the surface gravity  vanishes ($\kappa =0$). This is the reason that a non-zero surface gravity is a necessary condition for Hawking radiation emission \cite{visser03}. However of course the whole of this argument depends on the existence of the Killing vector field, and so will not be valid in a dynamical spacetime. Our argument above can be applied in that more general case. \\

In the next section we consider application of the redshift condition discussed here to the case of a cosmological black hole.

\section{Eikonal approximation for  cosmological black hole models}

The study of black holes in stationary and asymptotically flat
spacetimes has led to many remarkable insights. But, as we know,
our universe is not stationary and is in fact undergoing
cosmological expansion in presence of background radiation; that is the context in which we should consider gravitational collapse to form a black hole \cite{ellisetal14}.
There have been many papers constructing solutions of the Einstein equations
representing  a collapsing central mass in a cosmological context. Gluing of a
Schwarzchild manifold to an expanding FRW manifold is one such
attempt, made first by Einstein and Straus {\cite{Einstein Straus}. \\

Now, a widely used metric to describe gravitational collapse of
a spherically symmetric dust cloud is the so-called Lema\^{i}tre-Tolman-Bondi (LTB) metric {\cite{LTB}. Many people have looked at LTB models describing an overdense region in a cosmological background \cite{cosmological black hole}. Although there were  some attempts  to investigate  Hawking radiation from dynamical black holes through tunneling \cite{tunnelingbh}, no one has considered necessary conditions for this method in  presence of a matter flux such as that due to cosmological black body background radiation (CBR), which does indeed occur everywhere in the real universe.  \\

In this section we consider the essential features of black hole radiation emission for three models of cosmological black holes that take this effect into account: Oppenheimer-Snyder collapse, an LTB cosmological black hole, and a two-fluid model.

\subsection{Oppenheimer-Snyder collapse}
The Oppenheimer-Snyder solution consists of a dust filled Friedmann
model, joined across comoving spheres to a timelike hypersurface in the
Schwarzchild solution.  To make a cosmological black hole, we can match the exterior  Schwarzchild black hole to an internal spatially homogeneous FLRW universe (this is one kind of Einstein-Straus cosmological black hole). Since we want to consider the tunneling effect  near the horizon, it is sufficient to consider an  Oppenheimer-Snyder solution.
The metric inside (sign -) the collapsing dust in case of a flat 3-geometry is given by
\be
ds^2_-=-d\tau^2+a(\tau)^2(d\chi^2+\chi^2 d\Omega^2).
\ee
The Einstein equation shows that $a(\tau)$ satisfies
\be
\dot{a}^2=\frac{8\pi}{3}\rho a^2.
\ee
The equation of conservation,  $\nabla_\mu T^{\mu \nu} = 0$,  gives the density as $\rho(\tau)=\frac{\rho_0}{a^3}$. We assume that at an  initial time $\tau=\tau_0$ the scale factor is $a=1$, then we get
\begin{equation}
a(\tau) = (1-\frac{3}{2}\sqrt{\frac{8\pi \rho_0}{3}}(\tau-\tau_0))^{\frac{2}{3}}.
\end{equation} The star surface is located at  $\chi_0=constant$.
The metric outside (sign+) is given by
\be
ds^2_+=- f dt^2+ f^{-1} dR^2+R^2 d\Omega^2,
\ee
where $f=1-2M/R$.\\

 Since this coordinate system has a singularity at $R=2M$, we choose the Lema\^{i}tre coordinate \cite{novikov} to give a junction with the FLRW region which does not have a coordinate singularity (see Appendix~\ref{A} ).\\

Using the junction condition one can show
\be
M=\frac{4\pi}{3}\rho a^3\chi_0^3
\ee
which is constant. When the star radius is bigger than its Schwarzchild radius, there is no trapped surface, i.e $R>2M$. When the star falls into it's Schwarzchild radius, there will be two parts of the apparent horizon \cite{ellisetal14}. The first on the outside is the OMOTS (Outer marginally outer trapped 3-surface) which is the same as the Schwarzchild event horizon, and the second is the IMOTS (Inner  marginally outer trapped 3-surfaces) which is inside the star and reaches the singularity at $R=0$.  The trapped surface is located at $R=2M$ which is given by
\be
\chi|_{AH} = \frac{2M}{a}.
\ee
If this black hole is embedded in the cosmological  CBR radiation, it will not make much difference to the spacetime curvature, so we can treat this radiation in a linear approximation as a propagating field on the LTB background making little difference to the fluid collapse. It will however change the location of the OMOTS surface, which will become spacelike because of the CBR influx, which continually falls into the black hole as discussed in \cite{ellis13,ellisetal14}.\\

\begin{figure}[h]
\begin{center}
\includegraphics[scale = 0.34]{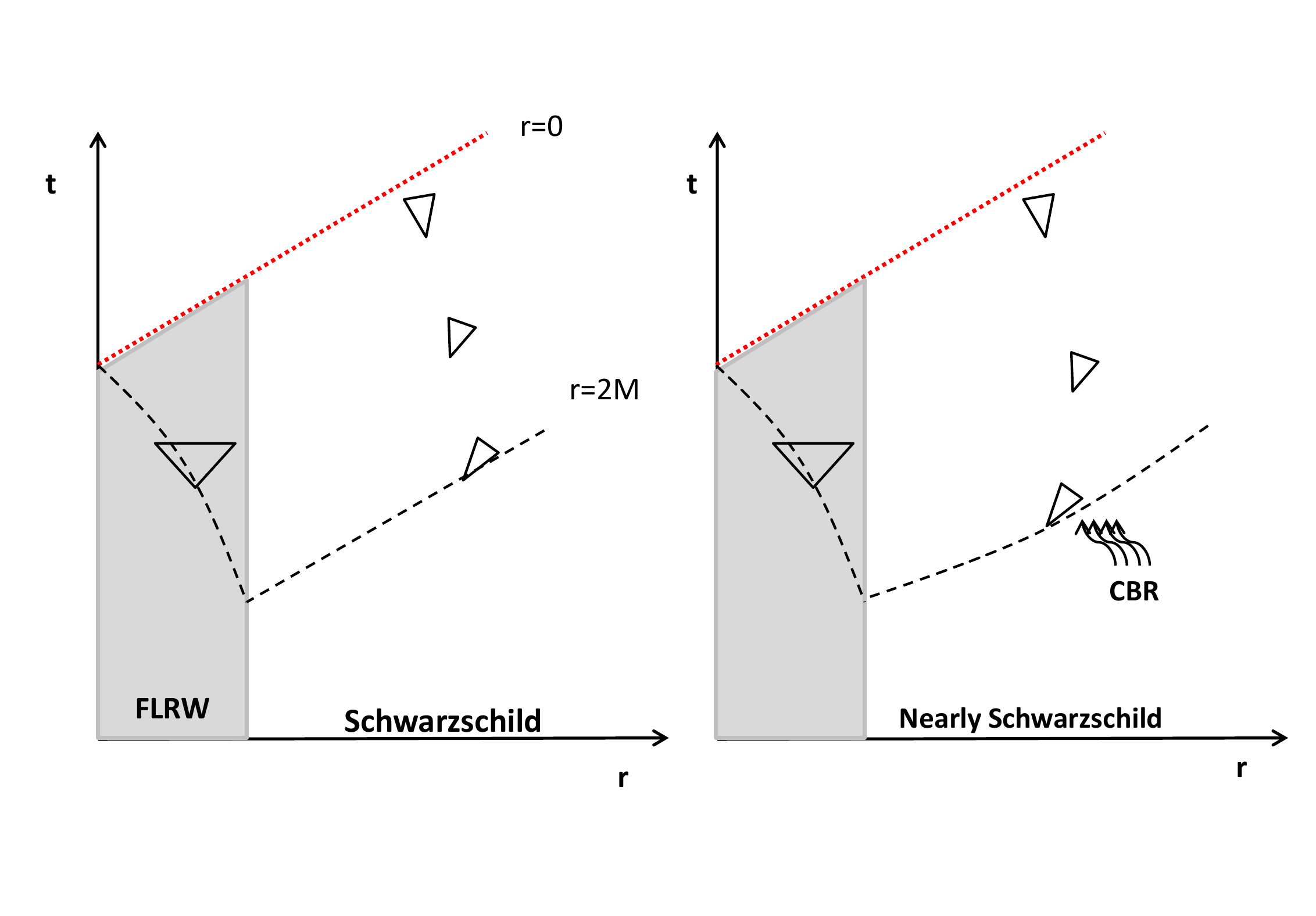}
\hspace*{10mm} \caption{ \label{oppenheimer}
 \textit{Oppenheimer-Snyder collapse in the presence of CBR and matter influx, in comoving coordinates. The left hand diagram is when there is  no infalling flux, and the right one is for the case of a matter or radiation influx that decays at late times.}}
\end{center}
\end{figure}
The radiation from the vicinity of the  OMOTS will be discussed in the next subsection, which is the more general case.
To know whether  this black hole emits radiation by tunneling from IMOTS, we have to check the geometric optic approximation for the ingoing and outgoing null geodesics which are emitted from 
a point near the IMOTS, see Fig.(\ref{oppenheimer}). Consider an  ingoing null geodesic which comes from a point (emitter) on or near  the IMOTS surface and arrives at an observer outside this surface. Since the space time is FLRW and the IMOTS is time like: $|\frac{d\tau}{d\chi}|_{null}<| \frac{d\tau}{d\chi}|_{IMOTS}$, the ingoing wave can pass between the two points without an infinite change in its phase.
 The outgoing null geodesic can exit from the timelike IMOTS to outside  without seeing any infinity (since the space time is FLRW).\\

 Hence,  the geometric optic approximation is not valid for ingoing or outgoing null geodesics emitted near the IMOTS.
As a result, the eikonal approximation condition \cite{visser03}  is not valid for this surface, and there is no black hole radiation from the  timelike IMOTS surface.
The Penrose diagram for the Oppenheimer-Snyder collapse with incoming CBR radiation is depicted in Fig.(\ref{PD_BH_CMBR}).
\begin{figure}[h]
\begin{center}
\includegraphics[scale = 0.65]{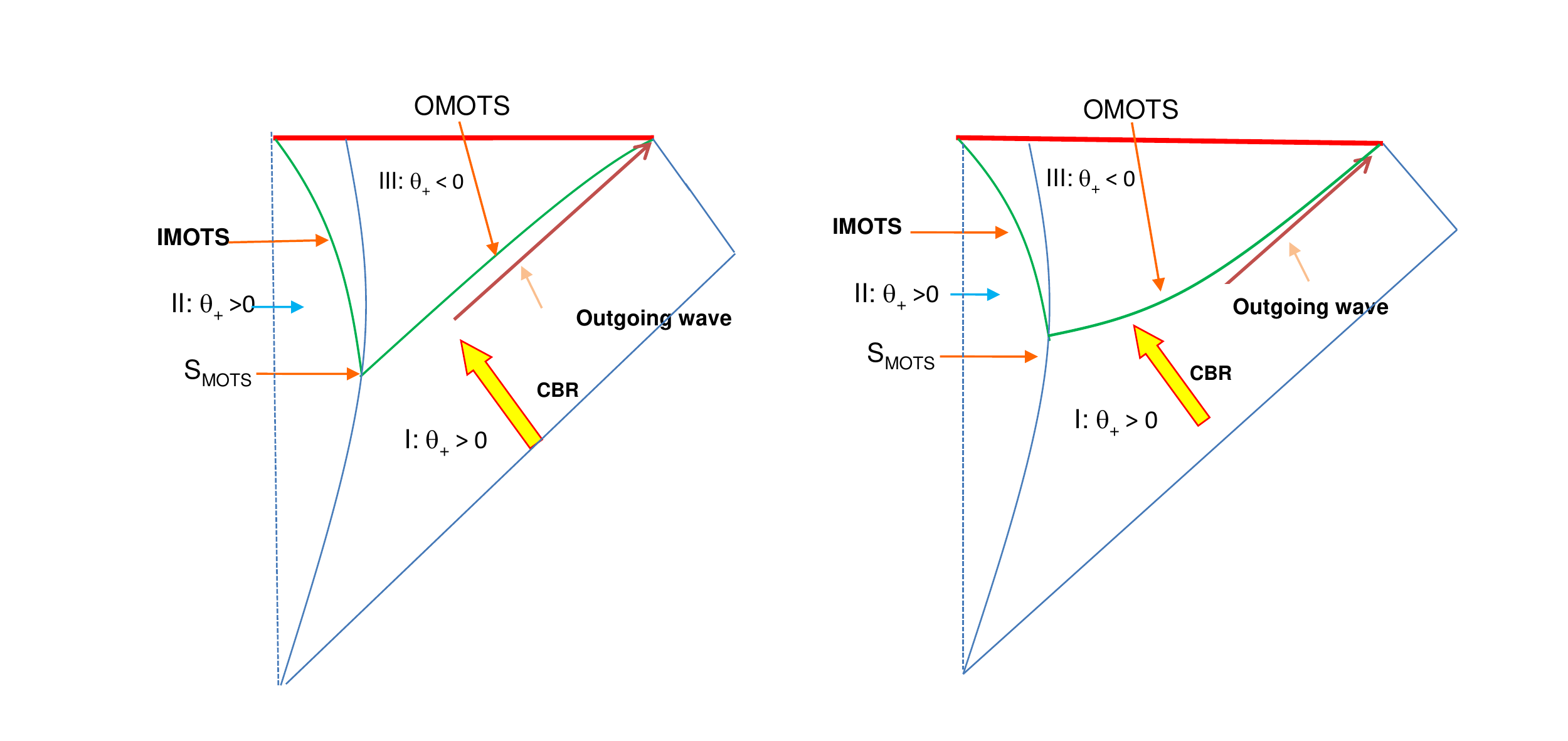}
\hspace*{10mm} \caption{ \label{PD_BH_CMBR}
\textit{Penrose diagrams of black hole formation with incoming CBR or matter flux. The left figure is the case of a constant infalling flux, and the right one shows the case of a decaying flux at late times. When the OMOTS surface is spacelike because of infalling radiation, as the outgoing null wave approaches that surface, the wave redshift seen for radiation emitted from close to this surface does not diverge. }}
\end{center}
\end{figure}

\subsection{LTB cosmological black hole}
 The LTB \cite{LTB} metric is a
spherically symmetric non-static solution of the Einstein equations
with a dust source. It can be written in synchronous coordinates as
\begin{equation}
 ds^{2}=-dt^{2}+\frac{R'^{2}}{1+f(r)}dr^{2}+R(t,r)^{2}d\Omega^{2}.
\end{equation}
and represents a pressure-less perfect fluid satisfying
\begin{equation}
\label{einstein}
\rho(r,t)=\frac{2M'(r)}{ R^{2}
R'},\hspace{.8cm}\dot{R}^{2}=f+\frac{2M}{R}+\frac{\Lambda}{3}R^2.
\end{equation}
Since this metric is dynamical and includes the general case of a  FLRW metric as well as inhomogeneous pressure free stars, the cosmological black hole can be modeled with this metric.
As discussed in \cite{firouzjaee-penn}, we can neglect the cosmological constant (considered as a dark energy candidate) for discussing local effects at the black hole such as black hole horizon dynamics.\\

In presence of the CBR and matter flux, we assume that most of the matter inside and around the black hole is dust causing a matter influx, and the CBR has a non-zero flux of radiation which falls in across the horizon and which can be treated as a linear perturbation, approximated as a propagating field on the LTB background and making little difference to the fluid collapse. It will however change the location of the OMOTS surface, which will become spacelike because of the CBR and matter influx \cite{ellis13,ellisetal14}. Thus the space time geometry can  effectively be  described by an LTB model embedded in the cosmological expanding background, in which the black horizon is dynamical owing to the non-zero incoming radiation flux.\\

To consider the geometric optics approximation on the apparent horizon let's look at null geodesics near this surface. Using the apparent horizon equation \begin{equation}
dR=2dM(r)=R'dr+\dot{r}dt|_{AH},
\end{equation}
the slope of the null geodesic vector tangent  to the apparent horizon tangent vector gives
\be
\frac{\frac{dt}{dr}|_{null}}{\frac{dt}{dr}|_{AH}}=\frac{R'}{R'-M'(r)}
\ee
To get this equation we have used the collapsing region condition $\dot{R}<0$ and equation (\ref{einstein}). This equation shows that if there were no matter flux into the black hole, $M'=0 \Leftrightarrow \rho =0$, the apparent horizon has to be null. If the matter flux is non-zero, the slope of the null geodesic is greater than that of the apparent horizon, which is spacelike \cite{ellisetal14}. In the case that there is no shell-crossing singularity, $R' \neq \infty $, and no other singularity on the  apparent horizon, the null geodesics slope is finite and non-zero.  \\

Generally the LTB apparent horizon can be either an IMOTS or OMOTS, with the nature of the latter depending on the matter falling into the black hole \cite{booth05}. In the case that the matter flux 
has positive energy density, the OMOTS apparent horizon is spacelike and lies inside the event horizon. \\

\textbf{No incoming CBR:} It has been shown \cite{firouzjaee-penn} that the redshift of the light emitted from the apparent horizon for a dust BH without incoming CBR is not generically infinite, but  in the case that the apparent horizon (OMOTS) is a slowly evolving horizon it is  infinite. As a result, Hawking radiation will occur from the OMOTS (slowly evolving horizon) for a dust cosmological black hole with no CBR.\\

\textbf{Incoming CBR:} Now we examine the redshift of the light from apparent horizons (both OMOTS and IMOTS) for a dust cosmological black hole in the presence of CBR radiation.  Since the CBR radiation gives a non-zero term for the flux of the matter, it causes the black hole area to grow \cite{ashtekar02}:
\begin{eqnarray}
\mathcal{F}= \mathcal{F}_{\rm matter} +\mathcal{F}_{\rm CBR} := \frac{1}{G}(M(r_2)-M(r_1))|_{AH},
\end{eqnarray}
which means $M' > 0$.
 For this black hole, the light redshift is \cite{firouzjaee-penn}:
\begin{eqnarray}
1+z &=& c_0 ~ exp \left(-\int_o^e \frac{\dot{R}'}{\sqrt{1+f}}dr \right) \nonumber\\
&=& c_0~ exp(\int_o^e
\frac{-1}{\sqrt{1+f}}\left(\frac{M'}{R\dot{R}}-\frac{MR'}{\dot{R}R^{2}}+\frac{f'}{2\dot{R}})
dr \right).
\label{ltbredshift}
\end{eqnarray}

\begin{figure}[h]
\begin{center}
\includegraphics[scale = 0.5]{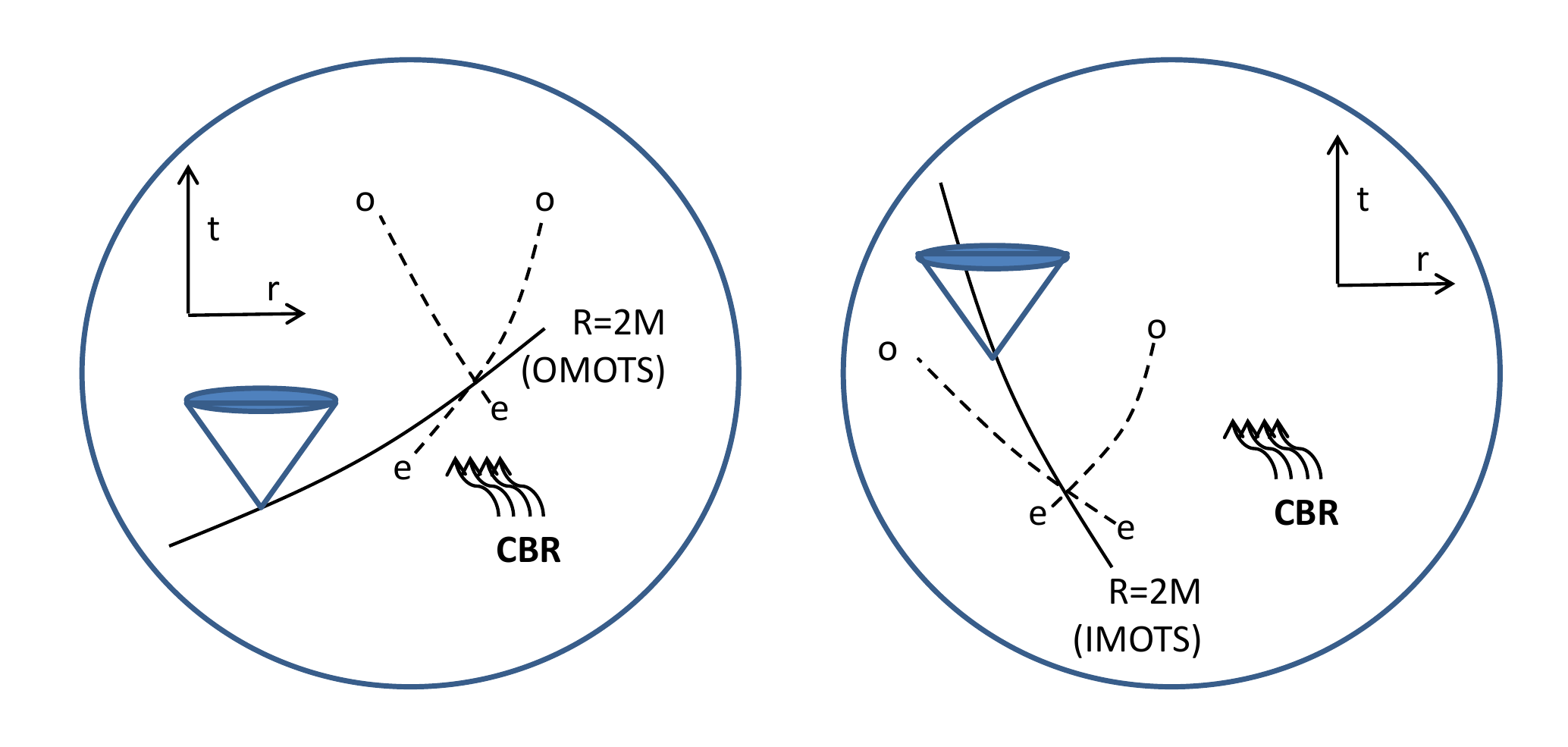}
\hspace*{10mm} \caption{ \label{ltbhorizon}
\textit{ The LTB causal  diagram near the IMOTS and OMOTS surfaces in the presence of CBR and matter influx, in comoving coordinates.}}
\end{center}
\end{figure}

Note that we calculate this redshift for a  comoving observer $u^\mu$ who is sitting at $r=constant$ with proper time $t$. We can choose another observer with 4-velocity $u'^\mu=\frac{dt}{dt'} u^\mu$, and we will see that the results will not change if there is no singularity in the second observer 4-velocity. As we know the apparent horizon is in the collapse region where $\dot{R}<0$ and $R$ is finite and non-zero there. With the assumption that there is no shell-crossing singularity and no infinity for density on the apparent horizon, we get  $R'$  and therefore $M'$  is finite.
As depicted in Fig.(\ref{ltbhorizon}), the ingoing and outgoing null geodesics which emerge from the space like apparent horizon (OMOTS) to a point inside it, will traverse a  finite $\Delta r$ because $\frac{dt}{dr}|_{null}$ is finite and non-zero. Similarly,  the ingoing and outgoing null geodesic which emerge from the time like apparent horizon (IMOTS) to a point outside or inside it, will traverse a finite $\Delta r$ because $\frac{dt}{dr}|_{null}$ is finite and non-zero as depicted in  Fig.(\ref{ltbhorizon}). Therefore, the exponential part of the equation (\ref{ltbredshift}) will be finite. Hence, the cosmological black hole apparent horizon in the presence of CBR radiation and matter flux does not satisfy the geometric optics approximation. Consequently, there is no Hawking radiation from either the IMOTS surface or the OMOTS surface in this black hole.
Note that, the causal structure in the  Fig.(\ref{ltbhorizon}) does not depend on the coordinates and the apparent horizon does not change this spacelike character, so all the above argument can be applied for other non-singular  coordinates.\\

This is in contrast to the Schwarzchild limit case, $M'=0$, where there is no incoming CBR and matter flux  and the apparent horizon tangent vector is a null vector (parallel to a null geodesic). Then the outgoing null geodesic traverses an infinite $ \Delta r \rightarrow \infty $ from the emitter point (on the horizon) to the observer point (this is the essential content of equation (2.16) in \cite{Haw73}). Therefore, in this case the apparent horizon is an infinite redshift surface and there is indeed black body radiation emission. \\

All the above discussion can be extended to  beyond the s-wave (spherical wave) case.  As discussed in \cite{visser03}, the near-horizon asymptotic behaviour, the phase pile-up, the continuation of the outgoing modes across horizon and Hawking temperature, are independent of having only an s-wave mode.\\

Even without these calculations, one can intuitively infer that since we do not have the adiabatic condition \cite{visser10} or late time apparent horizon (OMOTS) \cite{hajicek} for the dynamical black hole, there is no Hawking radiation. This is because null radiation emitted near the apparent horizon when the CBR and matter flux is significant would not be not tangent to the apparent horizon, which is spacelike, and so would arrive at infinity with a finite redshift ( Fig.(\ref{PD_BH_CMBR})).

\subsection{A two-fluid model}
The previous two models can be criticized because they do not model the effect on the spacetime of the CBR energy density. Here we consider a 2-fluid model representing the dynamic effects of both the matter and  the CBR radiation. As the latter is isotropic to very high accuracy in a FLRW model, we can represent it as a perfect fluid with 4-velocity $u^a_{CBR}$, density $\rho_{CBR}$, and pressure $p_{CBR} = \frac{1}{3}\rho_{CBR}.$ Thus the first fluid is ordinary matter fluid (non-zero inside the collapsing fluid, zero outside,) and the second one is the CBR radiation fluid. We assume the matter flux to be a perfect fluid, which can represent a realistic picture of gravitational collapse \cite{joshibook} and has the ability to model a cosmological black hole \cite{firouzjaee-penn}. \\

It can be shown that the  sum of the two perfect fluids is a fluid with anisotropic stress $\pi_{ab}$ and heat flow $q_a$ \cite{ellisbook}.
However we can set the heat flow to zero by choosing the timelike vector $u^a$ as the eigenvector of the Ricci tensor: then $q_{a}=0$. Because of the spherical symmetry of the problem, we can represent the anisotropic fluid pressure in this frame by two terms: a radial pressure $p_{r}$ and a tangential pressure $p_\theta$.  For slowly moving matter relative to the CBR, this effective fluid has
\be
\rho=\rho_m+\rho_{CBR}. 
\ee

The collapsing  fluid within a compact spherically symmetric
spacetime region will be described by the following metric in the comoving coordinates $(t,r,\theta,\varphi)$:
\begin{equation} \label{gltbm}
ds^{2}=-e^{2\nu(t,r)}dt^{2}+e^{2\psi(t,r)}dr^{2}+R(t,r)^{2}d\Omega^{2}.
\end{equation}
The energy momentum tensor for the fluid will have the
form
\begin{eqnarray}
 T^{t}_{t}=-\rho(t,r),~~T^{r}_{r}=p_{r}(t,r),~~T^{\theta}_{\theta}=
 T^{\varphi}_{\varphi}=p_{\theta}(t,r), 
\end{eqnarray}
with the weak energy condition satisfied:
\begin{equation}
\label{week}
 \rho\geq0~~\rho+p_{r}\geq0~~\rho+p_{\theta}\geq0.
\end{equation}
The Einstein equations give,

 \ba \label{gltbe2}
 \rho=\frac{2M'}{R^{2}R'}~,~~p_{r}=-\frac{2\dot{M}}{R^{2}\dot{R}},
\ea
\begin{equation} \label{pressure}
 \nu'=\frac{2(p_{\theta}-p_{r})}{\rho+p_{r}}\frac{R'}{R}-\frac{p'_{r}}{\rho+p_{r}},
\end{equation}
\begin{equation}
 -2\dot{R}'+R'\frac{\dot{G}}{G}+\dot{R}\frac{H'}{H}=0,
\end{equation}
where
\begin{equation}
G=e^{-2\psi}(R')^{2}~~,~~H=e^{-2\nu}(\dot{R})^{2},
\end{equation}
and $M$ is Misner-Sharp mass defined by

\ba \label{gltbe3} G-H=1-\frac{2M}{R}. \ea
As discussed in \cite{firouzjaee-penn}, the surface $R=2M$ is the apparent horizon for the collapsing region. In this case the matter flux which falls into the singularity is,
\begin{eqnarray}
\mathcal{F}= \mathcal{F}_{\rm matter} +\mathcal{F}_{\rm CBR} := \frac{1}{G}(M(t_2,r_2)-M(t_1,r_1))|_{AH},
\end{eqnarray}
As a result of non-zero CBR and matter flux, the density and the pressure are finite on the apparent horizon and then there are non-zero and finite values for $M' > 0$, $\dot{M} > 0$. No shell-focusing singularity, no shell-crossing singularity and the apparent horizon location in the collapsing region give  non-zero and finite values for $R$, $R'$ and $\dot{R}<0$ on the apparent horizon respectively.
The null geodesic slope on the apparent horizon is $\frac{dt}{dr}|_{null}=\pm\frac{R'}{|\dot{R}|}$ which according to the above discussion is finite for the outgoing and ingoing null geodesics. With some calculation one can get following equation on the horizon
\begin{equation}
\label{nullslop}
 \frac{\frac{dt}{dr}|_{null}}{\frac{dt}{dr}|_{AH}}=\left( \frac{1 - \frac{2 \dot{M}}{\dot{R}}}{1 - \frac{2 M'}{R'}} \right).
\end{equation}
In the vacuum (Schwarzchild limit) case $M'=0 \Leftrightarrow \rho =0$ and $\dot{M}=0 \Leftrightarrow p =0$, so the apparent horizon becomes a null surface. Using the weak energy condition (\ref{week}) we get $\frac{2 M'}{R'} \geq \frac{2 \dot{M}}{\dot{R}}$. The case of equality in the weak energy condition $\rho=-p$ is very special and non applicable to a realistic cosmological black hole, so we get
\be
\frac{2 M'}{R'} > \frac{2 \dot{M}}{\dot{R}}.
\ee
Using this equation in  (\ref{nullslop}) results in the dynamical case where the null geodesics  cross the apparent horizon (which is spacelike \cite{ellisetal14}),  and for the outgoing null geodesic we have
\begin{equation}
 \frac{\frac{dt}{dr}|_{null}}{\frac{dt}{dr}|_{AH}} > 1.
\end{equation}
In this way we can  calculate the light redshift from the apparent horizon of this metric. To this end, we need to calculate the affine parametrized radial null geodesic vector  $k^\mu$.  Without writing the details of the calculation, the result is that the  affinely parametrised null vector is
\be
k^\mu=c_0 e^{-\int (\dot{\nu}+\dot{\psi}+\nu' e^{\nu-\psi})\frac{dr}{e^{\nu-\psi}}}(1,\frac{\dot{R}}{R'},0,0)
\label{gltbredshift}
\ee
Considering a comoving observer with $u^\mu=(e^{-\nu},0,0,0)$, we can calculate the redshift of the null geodesics from the emitter point to the observer point. Using the geometric optics approximation, the question is whether the quantity $ e^{-\int (\dot{\nu}+\dot{\psi}+\nu' e^{\nu-\psi})dt}$ is infinite on the apparent horizon. Let us check all terms in this equation. The term $e^{\nu-\psi}=\frac{\dot{R}}{R'}$ on the horizon, which is finite.
We rewrite the equation (\ref{gltbe3}) as
\be
M=\frac{R}{2}(1-e^{-2\psi}(R')^{2}+e^{-2\nu}(\dot{R})^{2}).
\ee
The time derivative of this equation shows that if $\dot{\nu}$ and $\dot{\psi}$ are infinite, then $\dot{M}$ will be infinite, which as discussed above is an unphysical condition on the horizon.  Finally, with non-singularity of the pressure on the horizon, equation (\ref{pressure}) says that $\nu'$ is finite.
Overall, there is no infinity for the redshift of the null geodesic coming from the apparent horizon. Following the last section's discussion, there is no eikonal approximation for the null geodesics coming from apparent horizon (both OMOTS and IMOTS), and hence no emission of Hawking radiation.

\subsection{Necessary and sufficient conditions}
As pointed out by Visser \cite{visser03}, there are two other essential conditions  for having black hole radiation for a Lorentzian metric beside the eikonal approximation: the first is existence of an apparent horizon, and the second is non-zero surface gravity.\\

As regards the apparent horizon, its existence is necessary for defining the black hole boundary in dynamical collapse \cite{ashtekar02}. The Oppenheimer-Snyder (OS) model has an apparent horizon that consists of two parts, the IMOTS and the OMOTS \cite{ellisetal14}. For the LTB gravitational collapse model, it was shown in \cite{cosmological black hole, firouzjaee-penn} that an apparent horizon will form at $R=2 M$ after some time, and every LTB gravitational collapsing model has an apparent horizon. For the third (two-fluid) model of gravitational collapse, the scenario is different, because in such a model the gradient of the pressure acts as a repulsive force and can in principle prevent formation of the apparent horizon. However this depends on the details of the gravitational collapse such as the equation of state \cite{joshibook, moradi}, and for realistic equations of state a black hole will result if the initial mass of the collapsing object is large enough \cite{MTW,novikov}. For studying  black hole radiation in the general spherically symmetric case, the surface $R= 2M$ is again the black hole boundary, which is an apparent horizon.\\

The surface gravity definition for the first (OS) model is trivial. But the surface gravity definition for a general dynamical metric is not trivial. Having a covariant definition for quantities in thermodynamics  is very important because our physical quantities should not depend on the coordinates that we use to calculate. Fortunately, there is nice formula for the thermodynamic law in the spherically symmetric case \cite{hayward98} where the temperature is proportional to the surface gravity. The surface gravity definition of \cite{visser03} was rewritten in a covariant formalism in the spherically symmetric black hole case in \cite{tunnelingbh}. As shown in \cite{nielsen11}, and we calculate in equation (\ref{sgrav}) for the LTB metric,  the surface gravity is non-zero for spherically symmetric dynamical black holes.

\section{The future of the matter flux}

 As we have noted, the spacelike nature of apparent horizon (coming from the  in-falling matter or radiation flux) causes  light to pass it without having infinite redshift. Let us quantify this property. Consider the future-directed
 outgoing and ingoing null normals  $\l^a$ and  $n^a$ respectively at a point, and
the expansions $\theta_{\ell}$ and $\theta_n$  of the congruences of curves generated by
these vector fields.\\

Let $V^a$ be  tangential to $H$ (the MOTS hypersurface in Fig.(\ref{particle})), and orthogonal to the foliation by
marginally trapped surfaces. It is always possible to find a function $C$ and normalization
of $\ell^a$  such that $V^a = \ell^a - C n^a$. In addition,
the definition of $V^a$ implies that $\Lie_V \theta_\ell = 0$, which gives   an expression for $C$:
\be
C = \frac{\Lie_\ell \theta_\ell}{\Lie_n \theta_\ell} \, .
\label{defC}
\ee
When $C<0$ the  apparent horizon is an  IMOTS and  $C>0$ the apparent horizon is an OMOTS,  and if $C=0$ it becomes an event (isolated) horizon.
The value for the $C$ function is important because it shows the type of the black hole horizon. It shows if a MOTS surface is an OMOTS or IMOTS surface and whether it is timelike or spacelike (see Section 1.2 and Table 1 in\cite{ellis13}),  and specially, as discussed in the Appendix~\ref{C}, it is a criterion for where the adiabatic approximation is satisfied.\\

As discussed in  \cite{booth05}, in the case of perfect fluid collapse $C\propto (\rho+p)$ on the horizon. To see its behaviour let us look at the Friedmann  equation for the standard model of a $\Lambda CDM$ universe with FLRW metric
\be
(\frac{\dot{a}}{a})^2= \frac{8\pi G}{3} \rho + \frac{\Lambda}{3},
\ee
where energy conservation is
\begin{equation}
\dot{\rho}+3(\rho+p)\frac{\dot{a}}{a}=0.
\end{equation}
 The background matter density dilutes as  $\rho_m = \frac{\rho_m}{a^3}$. On the other hand, we know in the expanding background, only part of matter around the black hole can fall into it. For instance, for 
 a de Sitter universe representing the late evolution of a $\Lambda CDM$ model, there is a cosmological event horizon. Therefore, after some time the black hole devours all the available matter around itself and after that there is no matter flux which can make the apparent horizon a spacelike surface, i.e $C \rightarrow 0$. Note that this is a rough approximation because the matter around the black hole has pressure and the matter energy density increases when it gets  near the black hole horizon, so in a  more realistic model, $C$ is greater than this approximation and one needs a black hole simulation to find it.

\begin{figure}[h]
\begin{center}
\includegraphics[scale = 0.5]{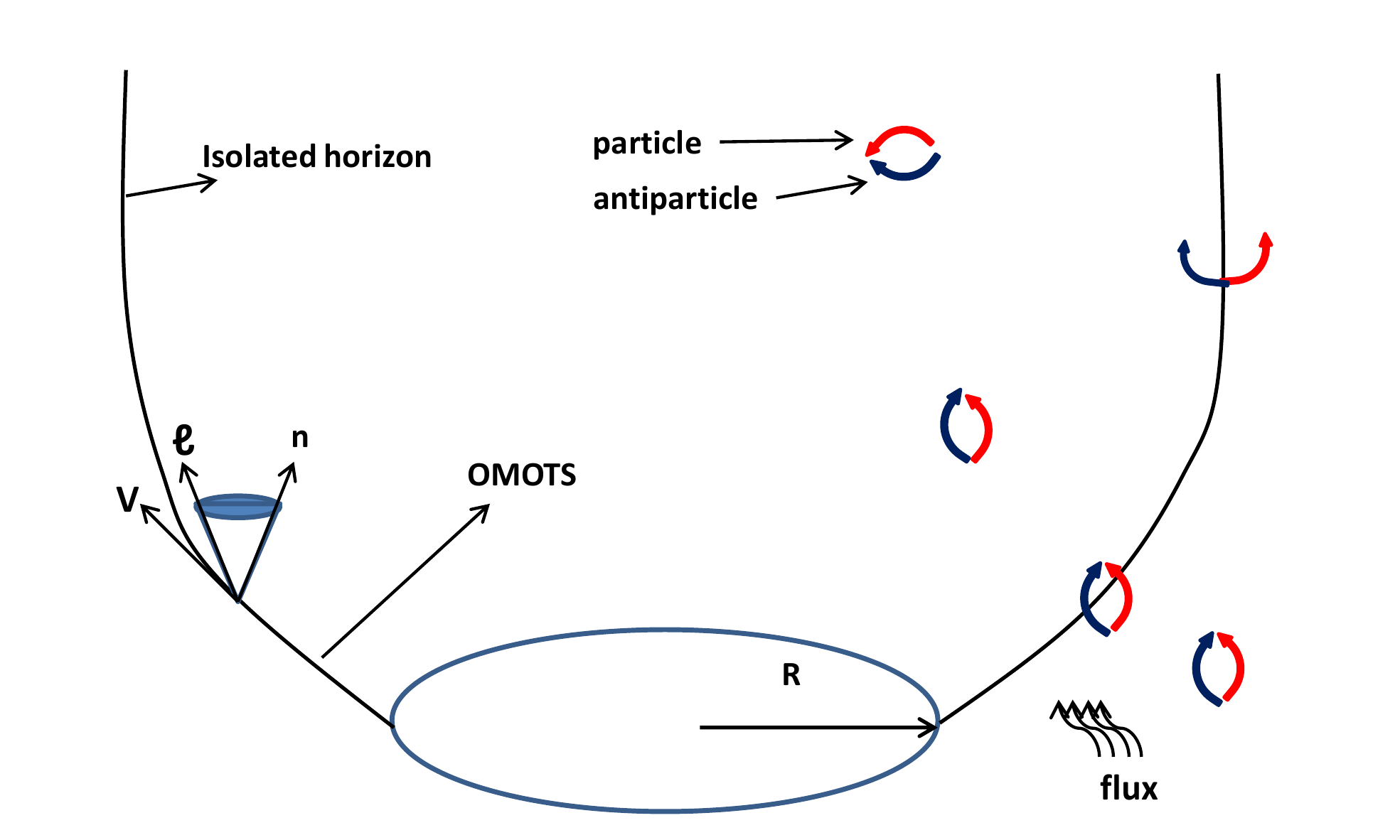}
\hspace*{10mm} \caption{ \label{particle}
 \textit{Particle creation and annihilation around the apparent horizon for the case if an influx that decays away at late times.
 The OMOTS surface starts at an $S_{MOTS}$ 2-sphere (bottom) and is first spacelike and dynamical; at this stage  virtual particle pairs can annihilate, and no radiation results. It then becomes first a slowly evolving horizon,  and then at very late times an isolated horizon.  At this stage virtual particle pairs can be separated by the OMOTS apparent horizon, and radiation results. Note that we did not include radiation backreaction in this Figure, which is adapted from arXiv:gr-qc/0308033.} }
\end{center}
\end{figure}

\subsection{ The CBR flux value }
In the last subsection we did not consider the CBR radiation. As discussed above, the $C$ function  (\ref{defC})  is a criterion for a MOTS surface being spacelike \cite{booth06,ellis13}.
To calculate the radiation flux we consider the CBR as a null fluid that moves inwards from infinity towards the center. In this case Booth \cite{booth05} has shown that the $C$ function is
\be \label{ccbr}
C= \frac{2G}{c^4} A_{AH}(\rho+p)
\ee
where $A_{AH}$ is apparent horizon area. At first sight, it seems that $C$ is very small for the current CBR energy density $\rho_{CBR} \simeq 10^{-14}~J/m^3$, but the basic point is that we have to calculate this energy density on the horizon, when CBR photons come from a large distance and get to the horizon. Therefore, the $C$ function is
\be
C= \frac{8G}{3c^4} A_{AH}~\rho_{CBR}~\frac{w_{AH}}{w_{background}}
\ee
where the $w$ is the CBR photon frequency for Kodama observers [8]. As shown in the Appendix~\ref{B},  for Kodama observers (which  reduce to static observers in the static case as discussed ref [8]), the CBR photon frequency becomes infinitely blue shifted when it gets to the apparent horizon.
  The infinity cannot have any meaning in classical physics. This is the trans-Planckian problem \cite{transplanck} which says that at the very onset of the formation of the trapping horizon, we must consider  semiclassical physics. We can roughly use the Planckian cutoff frequency to estimate the value of $w_{AH}$, which is $10^{43} H_z$. Let us present the general formula for a black hole with mass $M$ and at cosmological redshift time $1+Z_c = \frac{a_0}{a(t)}$ (the usual time scale in astronomy and cosmology). We know that the CBR background frequency is proportional $w_{0} \propto \frac{a_0}{a} \propto (1+Z_c)$ and the CBR energy changes  with time as $\rho \propto \frac{a_0}{a^4} =(1+Z_c)^4$. All in all, this leads us to this equation for the  CBR flux:
\be
 C \propto \frac{8G}{3c^4} 10^8 (\frac{M}{M_ \odot})^2~\rho_{CBR_0}~\frac{w_{AH}}{w_0} (1+Z_c)^3
\ee
If we use the present properties for CBR radiation $\rho_{CBR_0} \propto 10^{-14}$ and $w_0 \propto 10^{10}H_z$ we get
\be
 C \propto  10^{-16} (\frac{M}{M_\odot})^2~ (1+Z_c)^3
\ee
This shows that the CBR flux is important  for black holes with mass $M \gtrsim 10^7 M_\odot$ at the present time, because then $C$ is not negligible and so we then  do not have the adiabatic approximation, as shown in Appendix C.
But as we know, even if these black holes were radiating we would not see their temperature, because it will be $T<10^{-15} k $.\\

\textbf{Backreaction effects}  Assume that there is no infalling flux except the CBR flux, we consider now possible backreaction effects. Let us compare the  outgoing Hawking  radiation flux which decreases the black hole mass,  when that radiation is being emitted, with the infalling CBR flux, which cause  black holes to grow. If the CBR flux is greater than the Hawking radiation flux the OMOTS remains space like, and if the radiating flux is greater than the CBR flux, the OMOTS will eventually change to a timelike apparent horizon. Now the question is what is the magnitude of the $C$ function for the Hawking radiation flux? Similarly to   equation (\ref{ccbr}), the $C$ function becomes negative, giving a time like surface for the Hawking radiation-only  case:
\be
C= -\frac{G}{c^4} A_{AH}(\rho+p)_{HR}.
\ee
The black hole temperature is $T= 6 \times 10^{-8} \frac{M_ \odot}{M}~k$. Since the flux of the matter is proportional to the energy density \cite{ashtekar02}, we can write:
\be
\frac{\mathcal{F}_{\rm HR}}{\mathcal{F}_{\rm CBR}} = \frac{\rho+p|_{HR}}{\rho+p|_{CBR}}=\frac{C_{HR}}{C_{CBR}}
\ee
 We can compare this number with the $C$ function of CBR radiation which tells
\be
\frac{C_{HR}}{C_{CBR}}=- \frac{[6 \times 10^{-8} \frac{M_ \odot}{M}]^4}{[2.7~(1+Z_c)]^4}
\ee
For a solar mass black hole this expression  is $\frac{C_{HR}}{C_{CBR}}= 10^{-32}$  at the  present time, which is very small. For  black holes with mass $M > 10^{-8} M_ \odot$ the infalling CBR flux is greater than the radiation flux and OMOTS remains spacelike. Only for a black hole with mass $M \lesssim 10^{-8} M_ \odot$ are the two fluxes comparable and the black hole can have radiation which decreases the black hole area. This equation actually rules out  astrophysical black hole radiation backreaction effects that could make the OMOTS surface timelike, if there indeed is any Hawking radiation. Hence for such black holes the OMOTS surface is necessarily inside the classical event horizon. \\

However, a primordial black hole mass can be  less than this bound, then the CBR flux is negligible for them; but the matter flux value may still be significant, and its effect must be considered by modelling.

\subsection{The future of the CBR flux}

One may ask how long this scenario will remain true for a cosmological black hole when the universe expansion reduces the CBR flux effectively to zero in the far future. \\

Let us calculate the time evolution of the CBR flux.
For a universe dominated 
by radiation with $p = \rho/3$ one has
\begin{equation}
\frac{\rho_{CBR}(t_2)}{\rho_{CBR}(t_1)}=\frac{a(t_1)^4}{a(t_2)^4}.
\end{equation}
In the case of a matter dominated universe this term becomes
\be
\frac{\rho_{CBR}(t_2)}{\rho_{CBR}(t_1)}=\frac{(t_1)^{8/3}}{(t_2)^{8/3}}
\ee
while in the dark energy dominant era it is
\be
 \frac{\rho_{CBR}(t_2)}{\rho_{CBR}(t_1)}=e^{4H(t_1-t_2)}
\ee
where $H=\frac{\dot{a}}{a}$. The present value of the Hubble expansion rate is $H_0=3.241 \times 10^{-18} h s^{-1}$ where $h \simeq 0.72 \pm 0.1$.
 As discussed in last subsection, if we assume that the $C$ value decreases 2 orders of magnitude and OMOTS becomes nearly a null surface relative to the cosmological frame, this approximately means that the energy density decrease 2 orders of magnitude 
 (the $C$ function is key 
 to the validity of adiabatic condition (see Appendix~\ref{C}), hence we can use it to find approximately when the black hole radiation will start.) One can easily see that the times when the energy density of the CBR decreases to $ \frac{1}{100}$ of that at the present time in these two models
are $t_2 \simeq 2.44 \times 10^{18} s $ and $t_2 \simeq 7 \times 10^{17} s$ from the present time
respectively. Similar calculations can be done for the matter flux. Therefore, we can say that the  CBR and matter flux effect effectively cannot be diluted by the  cosmological expansion in the short term. In the other words, these calculations show that the cosmological expansion is not significant in turning  off the matter and CBR flux on  cosmologically short time scales. \\

However, leaving aside  strange behaviour such as dynamical dark energy, cosmological particle creation, and big crunch singularity models, this picture says that at very large times there is effectively no CBR flux. Although it never becomes exactly null or timelike (\cite{ellisetal14}, section VIII.2), the OMOTS eventually becomes an isolated horizon that is very close to the classical event horizon. Hawking Radiation emission could occur from that time on as depicted in Fig.(\ref{tpenrose}), because the approximations we have used above in the eikonal analysis would break down at that stage, with associated backreaction effects and possibly eventual black hole evaporation. \\

\begin{figure}[h]
\begin{center}
\includegraphics[scale = 0.7]{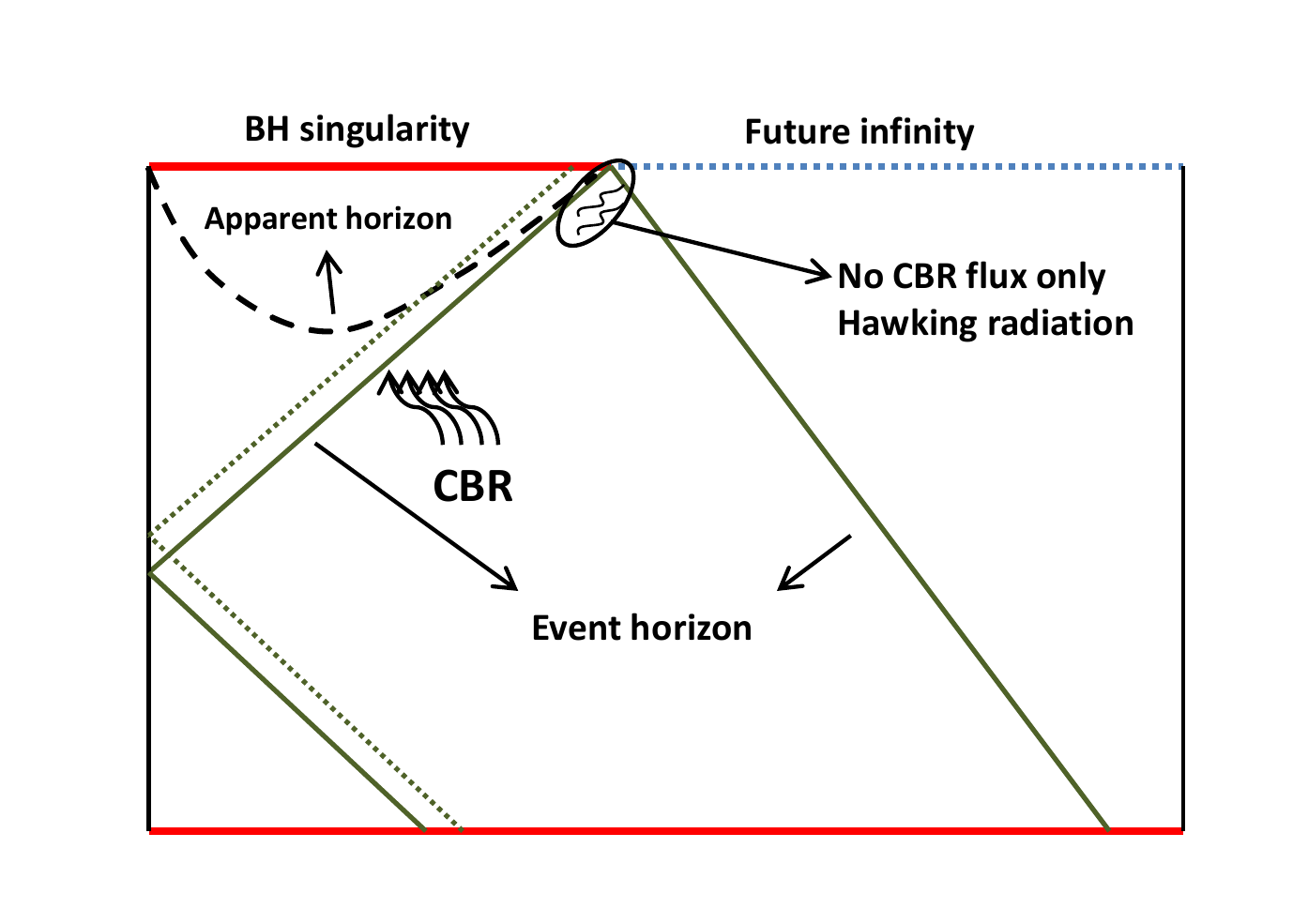}
\hspace*{10mm} \caption{ \label{tpenrose}
 \textit{Penrose diagram for a black hole in a $\Lambda CDM$ expanding universe, where there is  a matter and radiation influx, but it dies away to zero in the far future. Note that we do not represent back reaction effects in this picture.}}
\end{center}
\end{figure}

This is the key difference from the particle approximation and associated tunneling picture, because on that view, there would never be Hawking radiation emission, as the OMOTS surface would always be spacelike, so no Hawking Radiation backreaction effects or black hole evaporation would occur even in the very far future.

\section{Exponential approximation for the affine parameter}
This paper so far has been based on examining the Eikonal approximation and its implications. However the original calculation by Hawking \cite{Haw73} was based on the way that the affine parameters on past null infinity $\cal{I}_-$ and future null infinity $\cal{I}_+$ are related to each other by an exponential transformation, so an infinite range of the parameter $v$ on $\cal{I}_-$ corresponds to a finite range of the parameter $u$ on $\cal{I}_+$ (see Figure 4 of \cite{Haw73}).  Following this, in order to find the minimal condition for having thermal-like black hole radiation, Visser et al \cite{visser10} have  shown that any collapsing compact object (regardless of whether or not any type of horizon ever forms) will emit a slowly evolving Planckian flux of quanta, provided the adiabatic condition and exponential approximation hold along the null congruence which travels from past null infinity to future null infinity.
Having such a relation for  wave packets and  applying the standard Hawking calculation \cite{Haw73} to derive
time-dependent Bogoliubov coefficients, we can see this  gives us a Planckian spectrum with a time-dependent Hawking temperature.
Thus it was shown there that the exponential relation between the affine parameters on past null infinity, $v_{in}$, and future null infinity, $u_{out}$, is the necessary and sufficient condition for generating a Hawking flux.\\

In the case of black hole formation in a vacuum context, this leads to Hawking's picture. However in the case of a spacelike dynamical horizon as considered in this paper, the pile up of the u-parameter relative to the v-parameter does not reach infinity, because the OMOTs is spacelike (see  Fig.(\ref{bh_Ellis_javad_1})). The relevant null geodesics (for $u \rightarrow \infty $) cross the apparent horizon (the OMOTS surface) and do not reach future null infinity \cite{ellisetal14}. The null coordinate $u$ does not tend to infinity there, and hence surfaces of constant $w$ of the solution $p_w$ will not pile up at this surface as in the vacuum case considered by Hawking (\cite{Haw73}, just after Fig 4). \\

\begin{figure}[h]
\begin{center}
\includegraphics[scale = 0.7]{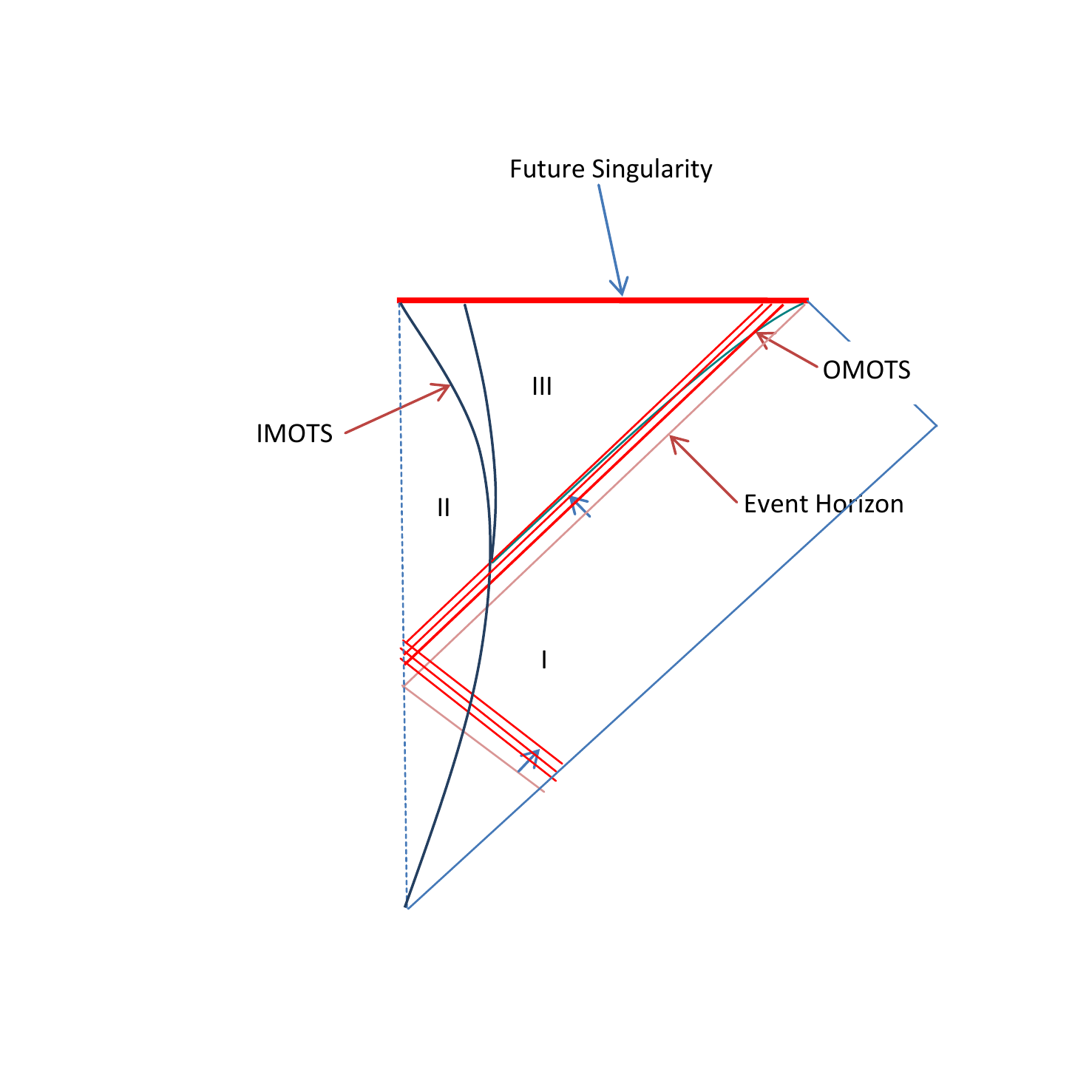}
\hspace*{10mm} \caption{ \label{bh_Ellis_javad_1}
\textit{Penrose diagram for black hole formation with incoming radiation that does not decay away. A spacelike OMOTS surface means the infinite affine parameter rescaling between ingoing and outgoing null geodesics that occurs in the pure vacuum case (\cite{Haw73}, Fig. 4) does not reach infinity: it is trapped by the singularity.}}
\end{center}
\end{figure}

Putting this in the LCDM context where future infinity is spacelike \cite{ellisetal14}, although there is a unique event horizon which is the inmost event horizon associated with the black hole, as the CBR becomes cooler and cooler, asymptotically the OMOTS will (in the preferred cosmological frame) tend to being null from being spacelike, but will never actually become null (Fig.(\ref{tpenrose})). It is then a delicate issue of how the limits work as to whether black hole radiation results from the null geodesics which pass near the slowly evolving horizon at very late times.\\

One might be concerned that neither the WKB approximation nor apparent horizons are necessary for the Hawking effect, see \cite{visser07} and \cite{Smerlak-13}. The point however is that we don't consider the most general features of spherical collapse and Hawking radiation. We look for the Hawking radiation phenomenon in the case of
cosmological black holes  which do indeed have apparent horizons  \cite{ellisetal14}.  We also have used the tunnelling approach \cite{parikh2000,pady2000,tunnelingbh,visser03},
for which the WKB approximation is a necessary assumption. As shown above, this means that we need a surface that has a very large redshift (not necessarily infinite) 
for the received light that is emitted from near the apparent horizon, and this leads to our results. In addition, to check the outcome, we have applied the adiabatic condition for cosmological black holes  in  Appendix~\ref{C}, which is another approach that confirms our results.

\section{Discussion and Conclusions}
\label{con}
There are several effects related to the vacuum in quantum field theory:  the zero point energy, Casimir effect, and dynamical Casimir effect (which is like a moving mirror). There are three more quantum vacuum effects that are due to a curved spacetime: metric quantum fluctuations, vacuum polarization, and Hawking particle creation \cite{novikov}. Particle creation results from the Hawking particle creation effect and dynamical Casimir effect (or moving mirror).  Basically, Hawking particle creation is thermal radiation due to the black hole horizon which is observed at a large distance. On the other hand, the dynamical Casimir effect is the production of particles and energy from an accelerated moving mirror (see \cite{brodag-book} for a comprehensive review). The question is how we can see this effect in a curved space time or in black hole collapse. In a flat space time, the dynamical Casimir effect appears due to a special boundary condition for wave solutions. We can include this effect in a curved space time by solving the general wave equation (including backscattering and boundary conditions) and finding the appropriate vacuum for it (which is like an initial condition). Hence, we can see traces of this effect in the expectation value of energy momentum tensor $ <T_{\mu \nu}>$ as vacuum polarization \cite{Dav76}.  \\

In our dynamical spherically symmetric space time where the apparent horizon is inside the event horizon, any virtual pair particle created  due to these effects cannot fall outside the event horizon, become real, and reach future infinity; however it can be seen locally in $ <T_{\mu \nu}>$ inside horizon. But for any general realistic model of star collapse, we must first solve the wave equation with suitable boundary conditions (the dynamical Casimir effect has dynamical boundary conditions) and second find the appropriate vacuum corresponding to the collapsing model (like the Unruh vacuum), and only then can we read the general particle creation occurring from the number density operator $< N=a^{\dagger}_k a_k >$ and $ <T_{\mu \nu}>$. It seems that in the analogue gravity models for Hawking radiation, we can see the dynamical Casimir effect \cite{analogue gravity}.  One can consider collapsing star models with special dynamical boundary condition to probe the dynamical Casimir effect for particle creation, which is beyond the scope of this paper. Here we just discuss the blackbody particle creation due to Hawking radiation.
\\

We have shown that when one uses either the tunneling approximation \cite{parikh2000}, \cite{tunnelingbh} or the eikonal approximation \cite{visser03}, one finds that turning on the CBR and matter flux turns off the Hawking radiation emission from the OMOTS surface, and there is also no such radiation emitted from the IMOTS surface. The key feature leading to this  result is shown in the contrast between the two cases in Fig.(\ref{oppenheimer}) and Fig.(\ref{tpenrose}):
\begin{itemize}
\item When we have the vacuum case outside the star, the OMOTS surface coincides with the null event horizon. Radiation emitted close enough to this surface reaches infinity with an unboundedly large redshift. It is this divergent redshift that is the reason Hawking radiation is emitted just outside the null OMOTS surface and escapes to infinity. The ultimate source of the divergent redshift is the infinite rescaling that takes place between the affine parameter and the group parameter on a bifurcate Killing horizon \cite{Boy69} (cf. equation (2.16) in \cite{Haw73}); hence it is a consequence of the static nature of the exterior vacuum solution, which allows this symmetry group.

\item When we take into account the infalling CBR radiation and matter flux, Fig.(\ref{tpenrose}), the OMOTS surface is spacelike and lies inside the null event horizon \cite{ellisetal14}. Outgoing rays reaching infinity from just inside its classical event horizon, or emitted inside that horizon but outside the OMOTS surface,  no longer experience this unbounded redshift; and the same applies to the IMOTS surface. That is the reason that  no Hawking radiation is emitted in this case.

 This is related to the fact that when the flux falls into the black hole, it's mass increases, hence this is no longer a quasi-static spacetime and there is no external Killing vector field with an associated bifurcate Killing horizon. The conclusion is reinforced  by the fact that there will be many other kinds of radiation and matter that will also fall into an astrophysical black hole, and increase its mass further. This is a self consistent approximation: as long as there is CBR and matter influx, the OMOTS surface will remain spacelike at all times because there  is no incoming negative density Hawking radiation that could make it timelike \cite{ellisetal14}.

\item As depicted in Fig.(\ref{tpenrose}), in the particle tunnelling scenario, if a pair of a particle and antiparticle are created near the dynamical horizon, both of them will fall into the singularity and annihilate each other, so no Hawking radiation will be emitted. However a particle created near an isolated horizon or slowly evolving horizon can reach  future infinity, and Hawking radiation will occur. On the other hand, since the geometric optics approximation does not generally held near the OMOTS, we cannot apply the particle interpretation for this surface.

\item At a late enough time in the very far future, the cosmological expansion decreases the CBR and matter density and black hole will devour all the available matter around itself. The black hole horizon then becomes first a slowly evolving horizon \cite{boothslow} and then an isolated horizon \cite{isolated} Fig.(\ref{tpenrose}). Black body radiation could then initiate at that stage, and possibly lead to black hole explosions at a later time.

\item As long as the matter flux into the black hole is not negligible there is no Hawking radiation, but the black hole radiation scenario is currently applied to every dynamical black hole. However, all black holes in the real universe are surrounded by different types of matter and radiation leading to substantial positive density energy influx for much of their life, and particularly in the very early universe.

\item Application of this constraint to  primordial black hole evaporation modelling may bring in a correction to their abundance in the cosmos. Specifically, primordial black holes are candidate progenitors of unidentified Gamma-Ray Bursts (GRBs) that are  supposed to  detect by the Fermi Gamma-ray Space Telescope observatory.
Their abundance might be lowered when the above considerations are taken into account.
\end{itemize}
This is all in accord with the discussions in \cite{ellis13,ellisetal14}, and leads to the conclusion that in a realistic cosmological context, a black hole forming from the collapse of a star in a universe permeated by CBR and matter will not emit Hawking radiation in the past or at the present, and so emission of such radiation from them, or evaporation of such black holes in an explosion, will not occur in the visible universe. To what degree this affects primordial black holes, or thefar future universe, will be very context dependent and will need detailed modelling. \\

One should contrast the above with the Parikh  and Wilczek tunnelling method (in standard general relativity) \cite{parikh2000}, which  cannot apply for particle production whenever the OMOTS surface is spacelike, because the whole concept of tunnelling only  makes sense for a timelike surface, where `inside' and `outside' are well defined concepts (see \cite{parikh2000}: paragraphs just after eqn. (8)). Therefore, the tunnelling picture is only applicable from the moment that the instantaneous Hawking radiation flux is greater than the matter flux and the black hole apparent horizon becomes a timelike surface.\\

Generally, in broader contexts, we cannot say that the tunnelling method requires the apparent horizon to be timelike, due to particle creation by tunnelling. One can apply this method even for spacelike universal horizons in Einstein-aether theory \cite{visser-13, prl-universal horizon} which has a non-standard causal structure. Note also that the adiabatic approach does not limit the horizon to be a timelike or spacelike surface. The point is that the adiabatic condition can be satisfied for an apparent horizon that becomes spacelike due to the CBR  flux. In addition, in our models, we quantify the WKB condition in the tunnelling method as the light redshift from the horizon, which is not a comprehensive approach. One can quantify this as the width of the horizon in order to calculate the lingering time in acoustic black holes or Lorentz-invariance violating models \cite{visser-13, width horizon}. 
\\

As mentioned above and emphasized in \cite{ellisetal14}, this work  should be extended in two significant ways:
\begin{itemize}
\item By calculating the expectation value of the stress tensor in this scenario (cf. \cite{BirDav84}),
\item By checking that radiation emission from within the collapsing fluid does not prevent horizon formation by back reaction of emitted Hawking Radiation, as has been suggested in \cite{Mer14}. That calculation would be altered in the cosmological context considered here, where \textit{inter alia} the Hartle-Hawking vacuum is not the appropriate vacuum state to use, and the spacelike nature of the OMOTS surface will modify the way modes propagating through the collapsing fluid \cite{Haw73,BirDav84} reach infinity.
\end{itemize}
However in terms of calculation methods that are used by various authors for determining the extent of Hawking radiation emission, the result given here seems conclusive.  Note that we do not claim to prove that no radiation \textit{at all} will be emitted. The break down of the adiabatic approximation implies only absence of a Planckian spectrum, not necessarily of any radiation. Indeed, a rapidly evolving apparent horizon would most probably lead to some form of particle production with typical frequencies excited of the order of the inverse timescale of the evolution of the metric (maybe linked to the accretion rate). This is a (geometrically induced) Dynamical Casimir effect, which per se can be far from thermal in character but not necessarily negligible (because the time dependence of the apparent horizon is supposed to be fast/non-adiabatic in the early universe).
\\ \\


{\bf Acknowledgments:}\\

We thank Malcolm Perry and Reza Mansouri for fruitful discussions,  and Ritu Goswami, Tim Clifton, David Jacobs and Matt Visser for helpful comments on an earlier version of this paper, as well as a  referee for useful comments. We thank the National Research Foundation (South Africa) and the University of Cape Town Research Fund for support.\\

\appendix

\section{Junction condition inside the horizon}\label{A}
To construct the Oppenheimer-Snyder model inside the Schwarzchild horizon, we need a coordinate system that does not have a singularity on the horizon.
 We choose the Lema\^{i}tre coordinate system \cite{novikov} which is similar to the FLRW and LTB comoving coordinates.
 Then
 \begin{eqnarray}
ds^2= -dt^2 + R'^2dr^2 + R(t,r) d\Omega^2
\end{eqnarray}
where
\begin{equation}
R= ( 2M)^{1/3} \left( \frac{3}{2} ( r-(t-t_0)) \right)^{2/3}.
\end{equation}
The singularity is  at $R=0$.
Since the induced metric must be the same on both sides of the star surface at $\chi_0$ , we get
\be
\label{match1}
R(\tau)|_{\chi_0} = a(\tau) \chi_0
\ee
and
\be
\label{match2}
(\frac{dt}{d\tau})^2|_{\chi_0} - R'^2(\frac{dr}{d\tau})^2|_{\chi_0} = 1
\ee
The unit normals ($n_{\mu} n^{\mu}=1$) to the star surface are $n_-^\mu=(0,a,0,0)$ and $n_{+\mu}=(-\frac{dr}{d\tau},\frac{dt}{d\tau},0,0)$.
Using these normal vectors to calculate the extrinsic curvature lead us to these equations:
\be
K_{-\theta}^{\theta} =K_{-\phi}^{\phi} = \frac{1}{a \chi}
\ee
and
\be
K_{+\theta}^{\theta} =K_{+\phi}^{\phi} = \frac{dr}{d\tau}\frac{\dot{R}}{R} +\frac{dt}{d\tau}\frac{1}{R'R}
\ee
Matching the two extrinsic curvatures on the star surface we get,
\be
\label{match3}
 (\frac{dr}{d\tau}\frac{\dot{R}}{R} +\frac{dt}{d\tau}\frac{1}{R'R})|_{\chi_0}= \frac{1}{a \chi_0}
\ee
The equations (\ref{match1}, \ref{match2}, \ref{match3}) give the evolution of the star surface in the two space times.\\

\section{Areal coordinate for dynamical metric}\label{B}
The $r$ coordinate which we used as a radial coordinate in the fluid is a  comoving coordinate,  where the time coordinate is proper time for this observer. In contrast to the stationary metric  in the vacuum, which has a preferred Killing observer, there is no preferred observer in the general dynamical metric. In the case that we want to calculate the redshift for a CBR photon  which comes from a large distance to the apparent horizon, we need better coordinates. In the spherically symmetric case, there is a well defined family of observers which can be attributed to the areal coordinate $R(t,r)$ \cite{kodama}, and which are Kodama observers. The  function $R(t,r)$ is a good candidate to describe the particle distance from center, and is called an areal coordinate (it is the angular diameter distance in cosmology).  By taking the
areal radius as a new coordinate and using the relation
$dR=R'dr+\dot{R}dt$ for the LTB metric one obtains
 \ba
 ds^{2}= (\frac{\dot{R}^{2}}{1+f}-1)dt^{2}+\frac{dR^{2}}{1+f}-\frac{2\dot{R}}{1+f}dR dt +R(t,r)^{2}d\Omega^{2}.
\label{ltbph} \ea
 The (t,R) coordinates are usually called areal coordinates. This LTB metric form is similar to the
Painlev$\acute{e}$ form of the Schwarzschild metric. In the
case of $f = 0$, the metric is the same as the Painlev$\acute{e}$ metric form.\\

The redshift (\ref{gltbredshift}) measured by Kodama observer with 4-velocity $K^{i}=\frac{\sqrt{1+f}}{(\sqrt{1-\frac{2m}{R}})R'}(R',-\dot{R})$ \cite{tunnelingbh}, can be written as
\begin{eqnarray}
1+z &=& c_0 ~\frac{\left(\frac{\sqrt{\frac{2m}{R}+f}-\sqrt{1+f}} {\sqrt{1-\frac{2m}{R}}}\right)_e}{\left(\frac{\sqrt{\frac{2m}{R}+f}-\sqrt{1+f}} {\sqrt{1-\frac{2m}{R}}}\right)_o}~ exp \left(-\int_o^e \frac{\dot{R}'}{\sqrt{1+f}}dr \right)
= c_0~ exp \left(-\int_o^e \pm  \frac{\dot{R}'}{\dot{R}+\sqrt{1+f}} dR \right).
\label{ltbredshiftR}
\end{eqnarray}
The $\pm$ refer to the outgoing and ingoing null geodesic respectively. Using the Einstein equation (\ref{einstein}), we get the following equation:
\begin{eqnarray}
1+z = c_0~\frac{\left(\frac{\sqrt{\frac{2m}{R}+f}-\sqrt{1+f}} {\sqrt{1-\frac{2m}{R}}}\right)_e}{\left(\frac{\sqrt{\frac{2m}{R}+f}-\sqrt{1+f}} {\sqrt{1-\frac{2m}{R}}}\right)_o}~ exp \left(-\int_o^e \pm  \frac{\dot{R}'}{- \sqrt{\frac{2M}{R}+f}+\sqrt{1+f}} dR \right).
\end{eqnarray}
The exponential part can be either $e^{-\infty}$ or $e^{+\infty}$ on the apparent horizon. For the ingoing null geodesic (with - sign), which comes from a large distance $e$ to the apparent horizon $o$, light becomes infinitely blue shifted:  $(1+z)\rightarrow 0$. Note that the coordinate $R$ (comoving for a Kodama observer), like the Schwarzchild areal coordinate given by the $R=constant$ surface, is a spatial coordinate outside the horizon ($R>2M$) but becomes a time coordinate inside the horizon. Therefore, this coordinate cannot be used for inside the apparent horizon $R<2M$.

\section{Adiabatic condition for dynamical metric}\label{C}
An equivalent statement to  the eikonal approximation is an adiabatic condition \cite{visser03, visser10}. This condition says that the need for slow evolution of the geometry is hidden in the approximation used to write the modes as $e^{iwt}$ multiplied by a position-dependent fact
or. This makes
sense only if the geometry is quasi-static on the timescale set by $w$.  This condition can be used as a criterion for having Hawking radiation \cite{nielsen12}. Since this condition is an essential condition for black hole radiation, we want to examine it for a dynamical metric in the Painlev$\acute{e}$-Gullstrand coordinates,
\be
ds^2=-(c(t,r)^2-v(t,r)^2)~dt^2-2 v dt~dR +dR^2+R^2 d\Omega^2
\ee
The adiabatic condition says that the peak in the Planck spectrum is meaningful when
\be
k T\approx w_{peak} \gg max\lbrace|\dot{c}/c|,|\dot{v}/v|\rbrace \label{adiabatic}
\ee
Now we examine this condition for the LTB metric (\ref{ltbph}) in this coordinate. The left hand side of this equation is proportional to the surface gravity \cite{tunnelingbh} which is finite. Without giving the details of the calculation for the  $v=\frac{\dot{R}}{1+f}$ function, we get

\be
\kappa_{AH} \gg |\dot{v}/v|= \frac{M'}{\sqrt{1+f}RR'}
\ee
Using the surface gravity for the LTB metric \cite{tunnelingbh}
\ba
\kappa_H=\frac{1}{2}g^{\mu \nu} \nabla_\mu \nabla_\nu R=\frac{1}{2R}-\frac{M'}{2RR'},\label{sgrav}
\ea
and the $C$ function for LTB \cite{firouzjaee-penn}, $C=2\sqrt{1+f}\frac{M'}{R'-M'}$, one gets
\begin{equation}
\label{ccc}
1 \gg C .
\end{equation}
Therefore, to examine the adiabatic condition for the LTB metric, it is sufficient to check  the $C$ function.\\

 \textbf{Conclusion:} Surprisingly, only in the case of a slowly evolving horizon obeying (\ref{ccc}) can a LTB dynamical black hole radiate. As a result, the adiabatic condition is also satisfied for an   isolated horizon where $C=0 \Leftrightarrow M'=0$.
\\

As another example, assume that we have ingoing radiation falling into the black hole. The suitable metric for this case is Vaidya ingoing form:
\be
ds^2=-(1-\frac{2M(v)}{r})dv^2+2dvdr+r^2 d\Omega^2,
\ee
where the apparent horizon is located at  $r=2M$. Using equation (\ref{sgrav}), we get $\kappa_H=\frac{1}{4M(v)}$ for the surface gravity. Hence, the adiabatic condition (\ref{adiabatic}) says
\be
1 \gg \frac{dM(v)}{dv}.
\ee
This means that the incoming matter flux must be small to satisfy the adiabatic condition. On the other hand, the $C$ function for a Vaidya metric that has a slowly evolving horizon  is \cite{booth06}
\be
1 \gg C=\frac{dM(v)}{dv}.
\ee
As a result, similarly to the LTB case, the adiabatic condition allows Hawking radiation for a Vaidya black hole that has a slowly evolving horizon or an isolated horizon. 


\end{document}